# Full-channel wavefront manipulation of surface waves with chirality-assisted geometric-phase metasurface

Shiqing Li[1,a,*], Min Kang[1,a], Jianru Li[3,4,a], Yueyi Yuan[2,a], Cong Liu[2], Xiaolong Liu[3], Juan Deng[1], Hang Zhang[1], Jinhua Yan[1], Linfang Shen[1], Bo Yan[1], Kuang Zhang[2,*], Lei Zhou[4,5,*], and Shulin Sun[6,*]

[1]*Department of Applied Physics, Zhejiang University of Technology, Hangzhou 310023, China*

[2]*Department of Microwave Engineering, Harbin Institute of Technology, Harbin, China*

[3]*Department of Electronics and Nanoengineering, Aalto University, Aalto, Finland*

[4]*Shanghai Key Laboratory of Metasurfaces for Light Manipulation, Fudan University, Shanghai 200433, China*

[5]*State Key Laboratory of Surface Physics and Key Laboratory of Micro and Nano Photonic Structures (Ministry of Education), Department of Physics, Fudan University, Shanghai 200433, China*

[6]*Shanghai Engineering Research Centre of Ultra Precision Optical Manufacturing, Department of Optical Science and Engineering, College of Future Information Technology, Fudan University, Shanghai 200433, China*

**Abstract**

Owing to their localized field enhancement and subwavelength resolution, surface waves (SWs) offer broad application potential in communications, sensing, and photonics via on-chip wavefront manipulation. This makes multi-channel SW wavefront manipulation highly desirable. However, conventional metasurfaces for SW wavefront shaping, relying on geometric and propagation phase mechanisms, typically exhibit similar functionalities for co- or cross-polarized output channels under different circularly polarized (CP) incidences, thereby limiting the development of high-capacity on-chip integrated devices. Here, by introducing the chirality-assisted phase as an additional phase control mechanism, we effectively decouple both co- and cross-polarized output channels, enabling independent SW wavefront shaping in four distinct channels. We numerically and experimentally demonstrate two metasurfaces in the microwave range: a four-channel SW meta-deflector and a four-channel SW metadevice that simultaneously produces a focused SW beam, a SW Bessel beam, and two deflected SW beams in different directions. Therefore, chirality-assisted geometric-phase metasurfaces provide a versatile platform for multi-channel SW wavefront engineering, offering significant potential for high-capacity on-chip communication and integrated photonic systems.

[a] Shiqing Li, Min Kang, Jianru Li, and Yueyi Yuan are co-first authors.

*Corresponding authors: Shiqing Li, sql@zjut.edu.cn; Kuang Zhang, zhangkuang@hit.edu.cn; Lei Zhou, phzhou@fudan.edu.cn; Shulin Sun, sls@fudan.edu.cn.

## 1. Introduction

Surface waves (SWs) are electromagnetic (EM) eigen-modes bounded at metal/dielectric interfaces. Two extraordinary properties of SWs (i.e., subwavelength lateral resolution and strong field enhancement at interfaces) [1-3] offer them many attractive applications, such as on-chip communications, sensing, and imaging [4-9]. In recent years, the rapid advancement of new-generation information technologies has substantially increased the demand for large-capacity SW transmission and processing [10,11]. Conventional optical elements, including lenses, prisms, and metal gratings, have been proposed for either exciting SWs [12-15] or manipulating their wavefronts [16,17]. However, due to the inherent nature of these elements, these traditional devices inherently suffer from the issues of bulky size, low efficiency and restricted functionalities.

Metasurfaces [18-20], which can provide abrupt phase shifts at subwavelength scales, enable unprecedented control over EM wavefronts [21-30], including SWs, making them ideal candidates for on-chip SW wavefront engineering. By simply varying the in-plane orientation of the constituent meta-atoms, the geometric phase (i.e., Pancharatnam-Berry (PB) phase) metasurface has been conveniently adopted in the manipulation of circularly polarized (CP) waves [31-40]. Physically, PB phase based metasurfaces support four distinct channels by switching the handedness of CP input and output beams. The two cross-polarized channels convert LCP input to RCP output (L-R) or RCP to LCP (R-L), while the two co-polarized channels preserve the input handedness (L-L and R-R). The inherent symmetrical response of PB phase results in that the functionality exhibited by metasurfaces in L-R channel and R-L channel would be opposite. Therefore, although PB metasurfaces have been widely used to excite SWs, they typically generate SWs symmetrically on both sides under CP wave illumination of opposite handedness [38-40]. To decouple this functional opposition and thus increase the number of SW control channels, combining propagation and geometric phases has been proposed to independently modulate SWs through the two cross-polarized channels (L-R and R-L) [41-43]. However, the propagation phase along the fast and slow axes influences the co-polarized output field phase without handedness selectivity, leading to similar responses for the L-L and R-R output channels. Very recently, frequency has emerged as an extra degree of freedom for multifunctional SW wavefront control [44-48]. By leveraging the structural heterogeneity of the constituent meta-atoms, on-chip four-channel SW manipulation at two distinct frequencies was achieved [48]. Nevertheless, these implementations still utilized solely the cross-polarized channels, whereas the co-polarized channels remained largely unexploited. Although considerable advancement has been achieved in SW wavefront shaping within multiple channels, existing metasurfaces are incapable of shaping diverse SW wavefronts through all four output channels at a single frequency, which constrains the development of on-chip multifunctional integrated devices.

To overcome this issue, we proposed a novel metasurface design strategy to achieve on-chip SW wavefront manipulation across all these four CP channels at a single frequency (Fig. 1). The designed metasurface consists of a top-layer transmissive metadevice and a bottom-layer plasmonic guiding region. The top-layer transmissive metadevice is designed by employing a hybrid mechanism that incorporates propagation and geometric phases together with a newly introduced chirality-assisted phase response (circular dichroism), to achieve four independent phase profiles across all co- and cross-polarized output channels under CP illumination. The bottom-layer plasmonic guiding region serves to transform the transmitted EMs into SWs possessing distinct wavefronts and sustain their propagation. To validate the proposed strategy, two metasurfaces with distinct functionalities under CP illumination were designed and experimentally verified. Under orthogonal CP wave incidence, the first metasurface operated as a SW meta-deflector capable of exciting SWs into four different deflection angles, while the second simultaneously generated a focused SW beam, a SW Bessel beam, and two deflected SW beams propagating in different directions. This work provides an additional step on the prospective way towards on-chip multi-channel SW wavefront manipulation, paving the way for large capacity on-chip communications and integrated photonic systems.

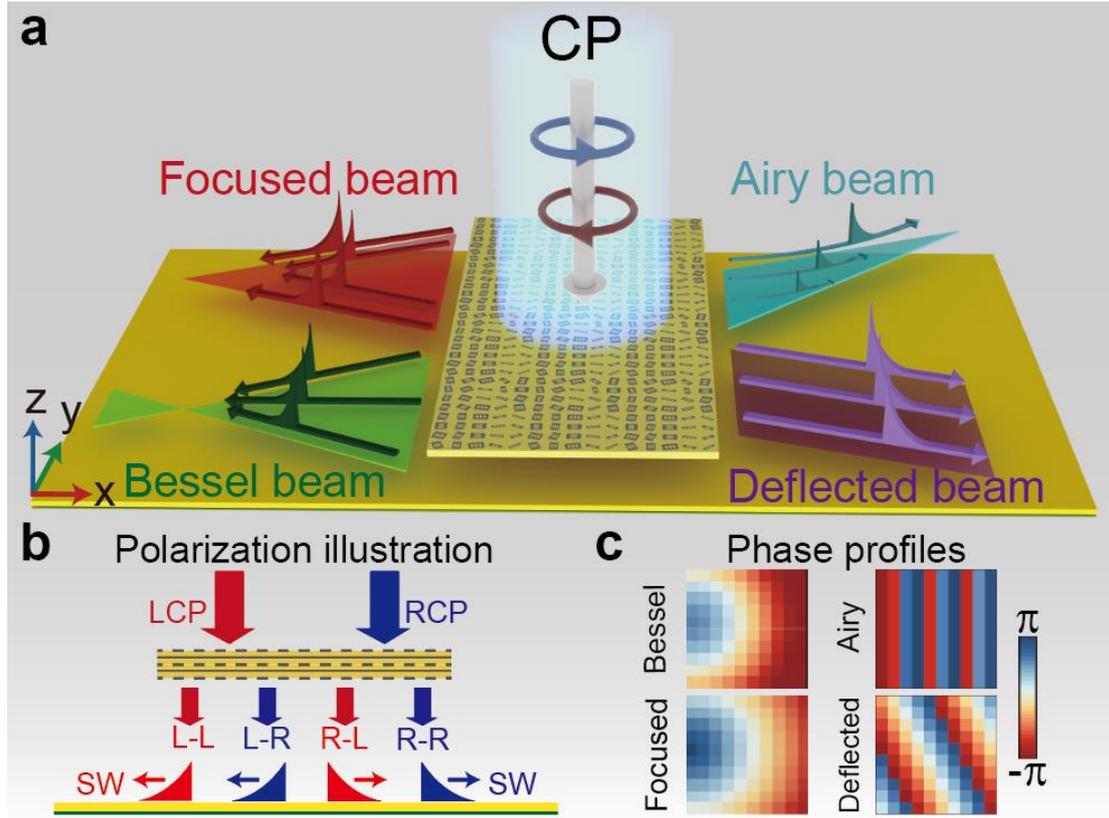

Fig. 1 Schematic principle of proposed metasurface for quad-channel wavefront manipulation of SWs. a Schematic diagram of the metasurface, upon incidence of CP waves on the top-layer metadevice, SWs with four types of wavefronts are generated on the bottom-layer plasmonic guiding region through L-L, L-R, R-L, and R-R channels. b The schematic side view of SW modulation through CP quad-channels. c The corresponding phase profiles for generating SWs with tailored wavefronts, including deflected, focused, Airy, and Bessel beams.

## 2. Principle and Design

### 2.1. General Scheme for Full-Channel SW Manipulation Metasurface

Here, we propose a general formalism for quad-channel SW manipulation through complete and separate phase manipulation of all four CP channels of metasurface. Arbitrary and independent wavefronts of SWs can be achieved by altering the CP states of input and output ends, as schematically illustrated in Fig. 1a. The proposed metadevice comprises a top-layer transmission-type metasurface and a bottom-layer plasmonic guiding region. The top-layer metasurface imparts distinct phase profiles across four CP transmission channels, whereas the plasmonic region transforms the resulting phase-modulated transmitted waves into SWs and facilitates their propagation, as depicted in the schematic of Fig 1b.

To achieve PW-to-SW conversion and thereby enable on-chip SW wavefront manipulation, we must engineer a phase profile such that not only does the phase gradient $\xi_x$ in the x-direction exceed the free-space wavevector k₀ ($\xi_x > k_0$), but the profile is also precisely designed along the y-direction. For example, to generate a focused surface-wave (SW) beam or a SW Bessel beam, the required phase profile can be expressed as $\Phi(x,y) = \Phi_0 - \xi_x x + k_{SW}(\sqrt{y^2 + F^2} - F)$ or $\Phi(x,y) = \Phi_0 - \xi_x x + \xi_y |y|$, respectively. Several target SW wavefronts and their corresponding required phase profiles are exhibited in Fig. 1c. Subsequently, we need to design a metadevice that achieves full phase modulation across four CP channels, with each channel encoded a distinct phase profile as shown in Fig. 1c.

To achieve distinct phase modulation in the four CP transmission channels, it is necessary to independently decouple the inherent correlations among the four output phase patterns φLL, φLR, φRL, and φRR (the first and second subscripts denote the input and output CP states, respectively, with L/R representing LCP/RCP). These phase patterns are determined by the phases of the four element transmission coefficients in the Jones matrix $T = \begin{bmatrix} t_{LL} & t_{LR} \\ t_{RL} & t_{RR} \end{bmatrix}$. The equivalent metasurface system is assumed to be passive, lossless, matched, and reciprocal. Therefore, the four CP transmission coefficients can be described in a linear basis as follows:

$$\begin{cases} t_{LL} = \frac{1}{2}[(t_{uu} + t_{vv}) + i \cdot (t_{uv} - t_{vu})], \\ t_{LR} = \frac{1}{2}[(t_{uu} - t_{vv}) - i \cdot (t_{uv} + t_{vu})] \cdot e^{i \cdot 2\theta}, \\ t_{RL} = \frac{1}{2}[(t_{uu} - t_{vv}) + i \cdot (t_{uv} + t_{vu})] \cdot e^{-i \cdot 2\theta}, \\ t_{RR} = \frac{1}{2}[(t_{uu} + t_{vv}) - i \cdot (t_{uv} - t_{vu})], \end{cases} \quad (1)$$

where $t_{uu} = |t_{uu}| \cdot e^{i\varphi_{uu}}$ and $t_{vv} = |t_{vv}| \cdot e^{i\varphi_{vv}}$ are the diagonal linear transmission coefficients, and $t_{uv} = |t_{uv}| \cdot e^{i\varphi_{uv}}$ and $t_{vu} = |t_{vu}| \cdot e^{i\varphi_{vu}}$ are the off-diagonal linear transmission coefficients. $\theta$ is the exterior rotation angle introduced by rotation matrix $M(\theta) = \begin{bmatrix} \cos\theta & \sin\theta \\ -\sin\theta & \cos\theta \end{bmatrix}$. Here, t_LL and t_RR are defined as copolarized transmission channels, which preserve the polarization state of input waves.

$t_{LR}$ and $t_{RL}$ denote cross-polarized channels that flip the output fields into opposite CP state.

According to Eq. (1), the full phase-modulation scheme can be constructed based on three phases, namely the propagation phase, the chirality-assisted phase, and the geometric phase, which constitute three degrees of freedom to decouple the inherent couplings between the circular-polarization (CP) transmission coefficients. These three phases determine, respectively, $\arg(\frac{1}{2}(t_{uu} \pm t_{vv}))$, $\arg(\frac{1}{2} \cdot i \cdot (t_{uv} \pm t_{vu}))$ and $\varphi_{PB} = \pm 2\theta$. Specifically, propagation phase modulation is used to set the initial phase profiles of the two diagonal and two off-diagonal transmission elements, $t_{LL}$ (= $t_{RR}$) and $t_{LR}$ (= $t_{RL}$), in the absence of chirality-assisted and geometric phase contributions. The chirality-assisted phase modulation is then introduced to decouple the interdependence between $t_{LL}$ and $t_{RR}$, while leaving $t_{LR}$ and $t_{RL}$ unchanged. Furthermore, geometric phase modulation can independently tailor the phase profiles of $t_{LR}$ and $t_{RL}$, without affecting $t_{LL}$ and $t_{RR}$.

Furthermore, in order to convert the transmitted EM waves carry four independent phase profiles into four SWs with distinct wavefronts and subsequently sustain their propagation, a bottom-layer plasmonic guiding region is required. Since natural metals exhibit excessively high conductivity to support surface plasmon modes at microwave frequencies, an artificial 'plasmonic metal' must be specifically designed.

### 2.2. Design and Simulation of Meta-Atom and Plasmonic Guiding Regions

The proposed metadevice consists of a transmission-type metasurface positioned a distance *d* above a plasmonic guiding region. The top-layer transmission-type metasurface is composed of multilayered meta-atoms engineered to implement all three phase modulation schemes, namely chirality-assisted phase, propagation phase, and geometric (PB) phase. The proposed meta-atom comprises five metallic layers and four dielectric substrates, with its geometric structure schematically shown in Fig. 2a. In our designed meta-atom, propagation phase is provided by tailoring the width $p_x$ and length $p_y$ of the rectangular patch layers, as illustrated in Fig. 2c, which also displays the corresponding phase profiles of all four output channels as a function of $p_x$. The results indicate that the phase responses are identical within the co-polarized channels $\varphi_{LL} =$

$\varphi_{RR}$ and within the cross-polarized channels $\varphi_{LR} = \varphi_{RL}$, while differing between the co- and cross-polarized groups ($\varphi_{LL} \neq \varphi_{LR}$, $\varphi_{RR} \neq \varphi_{RL}$). The phase profile tendencies as functions of both $p_x$ and $p_y$ are shown in Fig. S1 of SI. In order to decouple the intrinsic coherence between co-polarized output channels, the chirality-assisted phase is introduced via interior rotation of the meta-atom. As illustrated in Fig. 2d, this rotation is achieved by keeping the middle layer fixed while rotating the upper and lower layers by angles $\alpha_1$ and $\alpha_3$, respectively. Simulated results for a representative meta-atom with specific dimensions and varying $\alpha_1$ ($\alpha_3$ is fixed at $-\alpha_1$ in this special case) are also provided in the figure. It can be observed that the two co-polarized phase profiles exhibit distinct tendency with the interior rotation $\varphi_{LL} \neq \varphi_{RR}$ (derived and explained in Note 1 of SI). The two cross-polarized channels can be also decoupled by employing the geometric phase, which is imparted through an exterior rotation of the whole meta-atom structure by an angle $\alpha_2$, as shown in Fig. 3e. Figure 3e also presents the four output phase profiles of the representative meta-atom with specific dimensions and interior angle but various exterior angles, where two cross-polarized phases exhibit different tendency $\varphi_{LR} \neq \varphi_{RL}$ and two co-polarized phases are the same $\varphi_{LL} = \varphi_{RR}$. Finally, the meta-atom presented in Fig. 2a can be obtained by applying interior rotations $\alpha_1 = \theta_1 - \theta_2$ and $\alpha_3 = \theta_3 - \theta_2$, followed by an exterior rotation $\alpha_2 = \theta_2$, to a meta-atom of size $p_x \times p_y$.

Based on the theoretical and simulated results presented above, it is confirmed that the phase responses of all four CP channels ($\varphi_{LL}$, $\varphi_{LR}$, $\varphi_{RL}$, and $\varphi_{RR}$) can be independently tuned via the geometrical parameters ($p_x$, $p_y$, $\theta_1$, $\theta_2$, and $\theta_3$) of the proposed meta-atom. A meta-atom library that provides the required phase profiles for all CP conversion channels is established by sweeping the length $p_x$, width $p_y$, and interior angles $\alpha_1$ and $\alpha_3$, as detailed in Note 2 of SI. Given the target SW manipulation functions for the four channels, the required phase profiles are derived [41,42], which in turn allows appropriate meta-atoms to be selected from the library for metasurface construction. It should be noted that in this scheme, although independent phase manipulation is achieved, the amplitude responses are not involve d and the efficiency is partly sacrificed (see Note 4 and Fig. S4 in SI for details). Further independent amplitude modulation could potentially be realized by introducing additional degrees of freedom, such as material losses, active components, or even temporal variation.

The bottom-layer plasmonic metal is designed to convert EM waves from the top-layer metasurface with four independent phase profiles into four functionally distinct SWs and to support their subsequent propagation. Since natural metals are too conductive to support surface plasmon modes at microwave frequencies, the designed artificial 'plasmonic metal' incorporates a metallic ground plane covered with a 3.3 mm-thick dielectric layer ($\varepsilon_r = 3.5 + 0.01i$, as shown in the inset of Fig. 2b). Figure 2b depicts the FEM-simulated dispersion relation of the SW mode supported by such a system, which exhibits an eigen wave-vector $k_{SW} = 1.2k_0$ ($k_0$ being the free-space wave-vector) at the target frequency of 10 GHz.

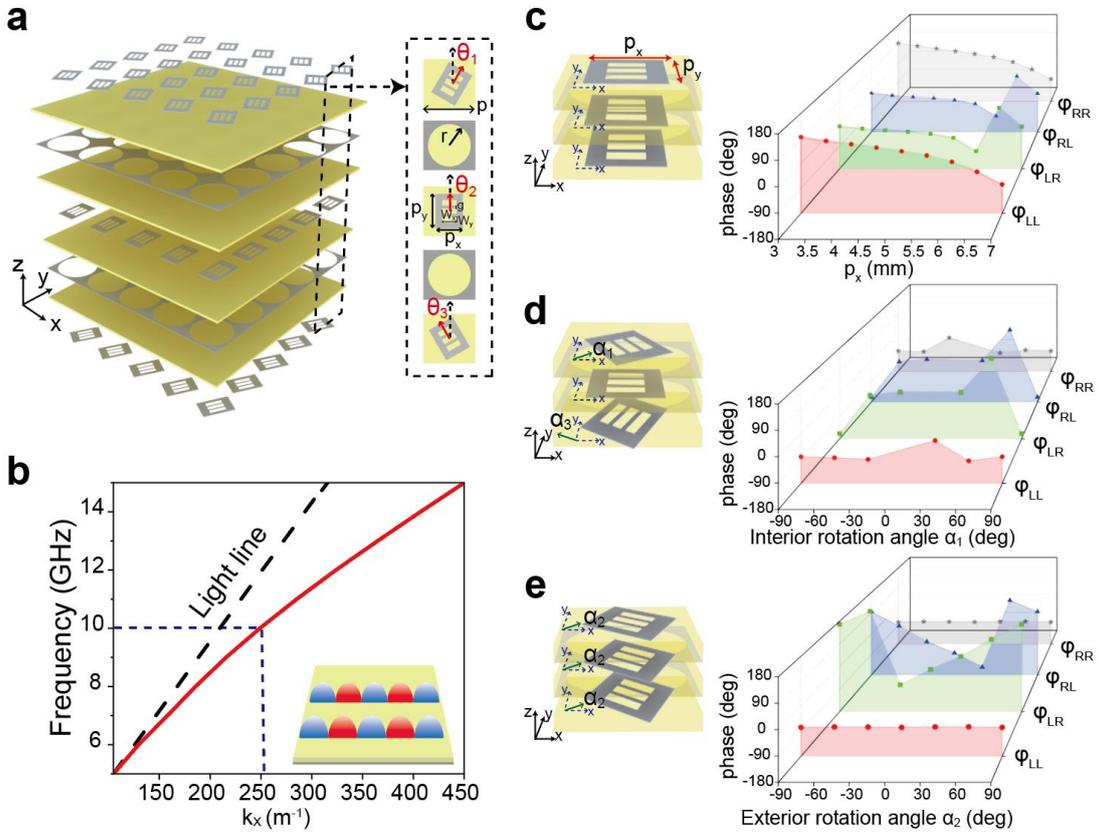

**Fig. 2 Simulation verification of the proposed meta-atom and bottom-layer plasmonic metal structure. a** Schematic of proposed meta-atom geometric structure, where periodicity is $p = 10$ mm, radius of circular aperture in the grid layer is $r = 4.5$ mm, thickness of each dielectric substrate is $h = 1$ mm, $p_x$ and $p_y$ are the length and width of metallic patch, $w_x$ and $w_y$ are the dimensions of rectangular gap in the metallic patch, $g$ is the distance between two adjacent gaps, $\theta_n$ (n = 1, 2, 3) denote the rotation angles of three patch layers. **b** The finite-element-method (FEM) simulated dispersion relation (red line) of the eigen SWs supported by the plasmonic metal as depicted in the inset. **c** 3D view of meta-atom with original state, and the output phase profiles of four channels against length $p_x$ of the patch layer for verifying propagation phase. **d** 3D view of meta-atom with only interior rotation, and the output phase profiles of four channels against interior rotation angle $\alpha_1$ for demonstrating chirality-assisted phase. **e** 3D view of meta-atom with only exterior rotation, and the output phase profiles of four channels against exterior rotation angle $\alpha_2$ for demonstrating geometric phase.

### 3. Results and Discussions

#### 3.1. Quad-Channel SW Meta-Deflector

After establishing the working principle and meta-atom library, we now demonstrate two metadevices for full-channel SW wavefront manipulation under orthogonal CP incidences. The first metadevice acts as a SW deflector that converts normally incident CP waves at the target frequency of 10 GHz into SWs and deflects them toward four independent directions, with its functional schematic illustrated in Fig. 3a. To this end, the top-layer metasurface is required to satisfy the following phase distributions [41]:

$$\begin{cases} \phi_{\text{LL}}(x,y) = \phi_0 - \xi_x x + \xi_y y \sin \vartheta_1 \\ \phi_{\text{RL}}(x,y) = \phi_0 + \xi_x x + \xi_y y \sin \vartheta_2 \\ \phi_{\text{LR}}(x,y) = \phi_0 + \xi_x x + \xi_y y \sin \vartheta_3 \\ \phi_{\text{RR}}(x,y) = \phi_0 - \xi_x x + \xi_y y \sin \vartheta_4 \end{cases} \quad (2)$$

where $\xi_x = 1.2k_0$, $\xi_y = 1.2k_0$, $\vartheta_1 = \vartheta_2 = 18°$, $\vartheta_3 = \vartheta_4 = -18°$ for the present case, and $x$ and $y$ denote the geometric center position of each meta-atom in the top-layer metasurface. This implies that the theoretical output tilting angles of the SWs are preset as rotations about the $z$-axis: $\vartheta_1 = 18°$, $\vartheta_3 = -18°$ for the L-L and L-R channels with respect to the $-x$ direction, and $\vartheta_2 = 18°$, $\vartheta_4 = -18°$ for the R-L and R-R channels with respect to the $+x$ direction. The strategy described in Sec. 2.2 guides the design and fabrication of our four-channel SW meta-deflector, as shown in Fig. 3c (see Fig. S2 in SI for details).

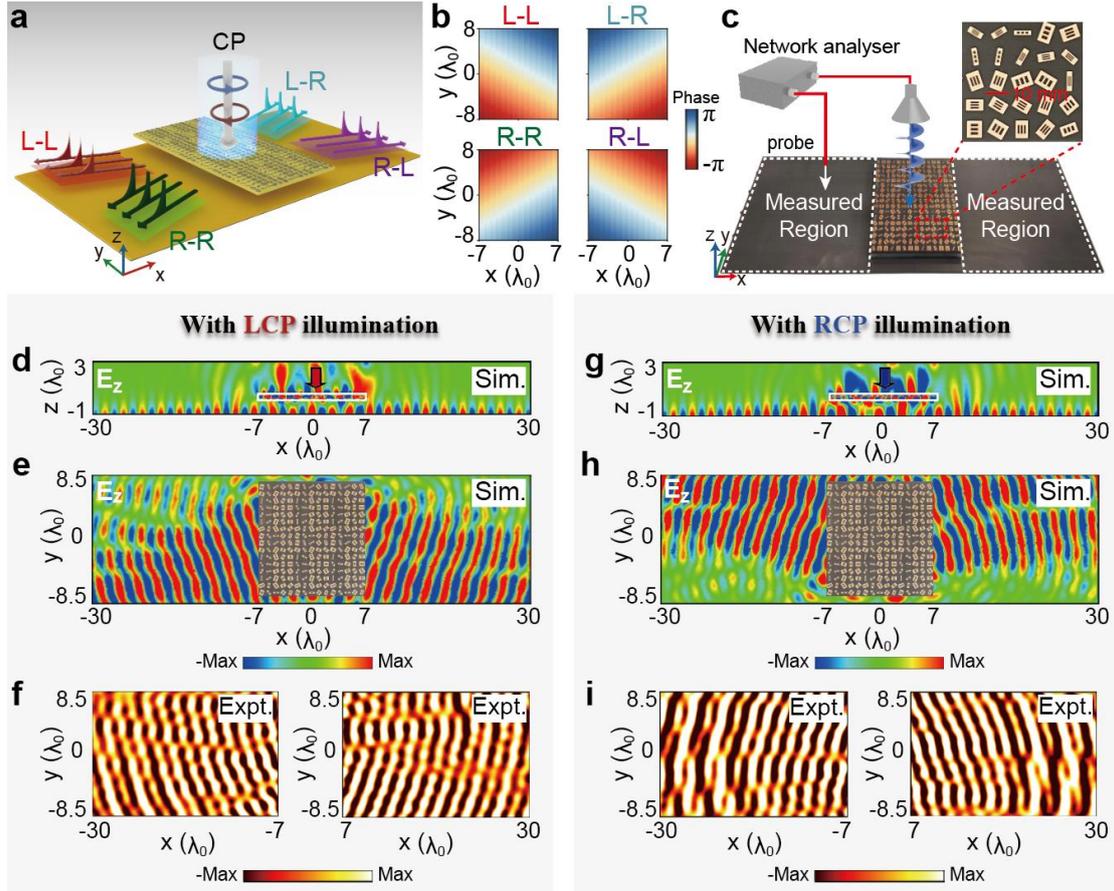

**Fig. 3 Numerical and experimental results validating the performances of the quad-channel SW meta-deflector. a** Schematic of proposed SW meta-deflector. **b** Simulated phase profiles of the meta-atoms within the designed metadevice for all four CP transmission channels. **c** Schematic of the experimental setup consisting of the fabricated top-layer metasurface placed at 10 mm above the bottom-layer plasmonic metal, a source horn, a monopole antenna, and a network analyzer. The monopole antenna is used to probe the SW field distribution on the plasmonic metal. **d-f** Under LCP wave incidence, the simulated near-field Re[$E_z$] patterns in the *xoz* (**d**) and *xoy* (**e**) planes, along with the corresponding measured results (**f**). **g-i** When flipping the incident state to RCP incidence, the simulated near-field Re[$E_z$] patterns in the *xoz* (**g**) and *xoy* (**h**) planes, along with the corresponding measured results (**i**). Here, the working frequency is 10 GHz.

We first employed FEM simulations to verify our theoretical predictions. Figure 3b presents the phase distribution $\varphi(x, y)$ of the transmitted waves from the top-layer metasurface for all four channels, which is in excellent agreement with Eq. (2), confirming that the desired phase profiles are successfully realized. Figure 3d, e, g, h depict the simulated Re[$E_z$] field distributions on the planes $y$ = 0 mm and $z$ = 1 mm at 10 GHz under orthogonal CP incidences. Under LCP incidence, the space wave is efficiently converted into SWs propagating along the bottom-layer plasmonic guiding region, forming two deflected SW beams with deflection angles of +19° (left side) and −17.8° (right side), as shown in Fig. 3d, e. When the helicity of the incident wave is flipped to RCP, the result is nearly symmetric to the case under LCP incidence, with SW deflection angles of −18.2° (left side) and +18.4° (right side), respectively (Fig. 3g, h). To quantitatively evaluate the conversion performance, we calculated the SW conversion efficiency, defined as the ratio of the power carried by the excited SWs to the total incident power [41,42,49]. The computed conversion efficiencies are 47.4% (including 24.8% for L-L channel and 22.6% for L-R channel) and 46.0% (including 19.5% for R-L channel and 26.5% for R-R channel) under LCP and RCP illumination at 10 GHz, respectively. These efficiencies are relatively modest, which can be mainly attributed to two factors: (1) the transmission efficiencies of the individual meta-atoms are inherently limited; (2) the designed plasmonic metal supports only transverse-magnetic (TM) polarized SWs, whereas the incident CP wave contains both TM and transverse-electric (TE) components [40]. Nevertheless, these efficiencies are already well above 45%, which is sufficiently high for practical applications. Further performance improvement can be achieved by replacing the top-layer metallic meta-atoms with dielectric ones and adopting a more sophisticated plasmonic guiding region that supports both TE- and TM-polarized SWs. Finally, we note that the distance $d$ between the top-layer transmission-type metasurface and the bottom-layer plasmonic metal has an important influence on the SW efficiency, and the highest efficiency can be obtained when $d$ is fixed at a specific value $d_c$ = 10 mm. (see Note 5 and Fig. S5 in SI for details). Consequently, the proposed metasurface enables efficient excitation of SWs at the target frequency under orthogonal CP incidences, providing a promising platform for on-chip multi-channel SW wavefront manipulation.

We finally fabricated the metadevice and adopted the near-field scanning technique to experimentally characterize its performance (see Note 6 and Fig. S6 in SI for details). As illustrated in Fig. 3c, the top-layer metasurface was illuminated by normally incident CP waves emitted from a horn antenna, and a monopole antenna was used to map the local Re[$E_z$] field distribution on a plane 1 mm above the bottom-layer plasmonic metal. Both the monopole antenna and the horn antenna were connected to a vector network analyzer (Agilent E8362C PNA). Figures 3f and 3i show the measured Re[$E_z$] patterns at the designed operating frequency of 10 GHz when the metadevice is illuminated by CP waves with opposite helicity. The measured patterns clearly demonstrate the expected full-channel SW manipulation capability of the metadevice. The measured SW deflection angles are −18.8°, +18.2°, +19°, and −18.5° for the L-L, L-R, R-L, and R-R channels, respectively, which are in good agreement with both theoretical predictions and full-wave simulations.

**3.2. Quad-Channel SW Manipulation with Distinct Functionalities**

To further demonstrate the full control of the four CP conversion channels for generating SWs with distinct functionalities, we design a metadevice that can produce a focused SW, a Bessel-beam SW, and two deflected SWs at 10 GHz under orthogonal CP incidences, as illustrated in Fig. 4a. To this end, the top-layer metasurface is required to satisfy the following phase distributions:

$$\begin{cases} \phi_{LL}(x,y) = \phi_0 - \xi_x x - \xi_y \left(\sqrt{y^2 + F^2} - F\right) \\ \phi_{RL}(x,y) = \phi_0 + \xi_x x + \xi_y y \sin \vartheta_5 \\ \phi_{LR}(x,y) = \phi_0 + \xi_x x + \xi_y y \sin \vartheta_6 \\ \phi_{RR}(x,y) = \phi_0 - \xi_x x - \xi_1 |y| \end{cases}, \quad (3)$$

where $\xi_x = 1.2k_0$, $\xi_y = 1.2k_0$, $\xi_1 = 0.3k_0$, $F = 185$ mm, $\vartheta_5 = -18°$, $\vartheta_6 = 18°$ for the present case, and $x$ and $y$ denote the geometric center position of each meta-atom in the top-layer metasurface.

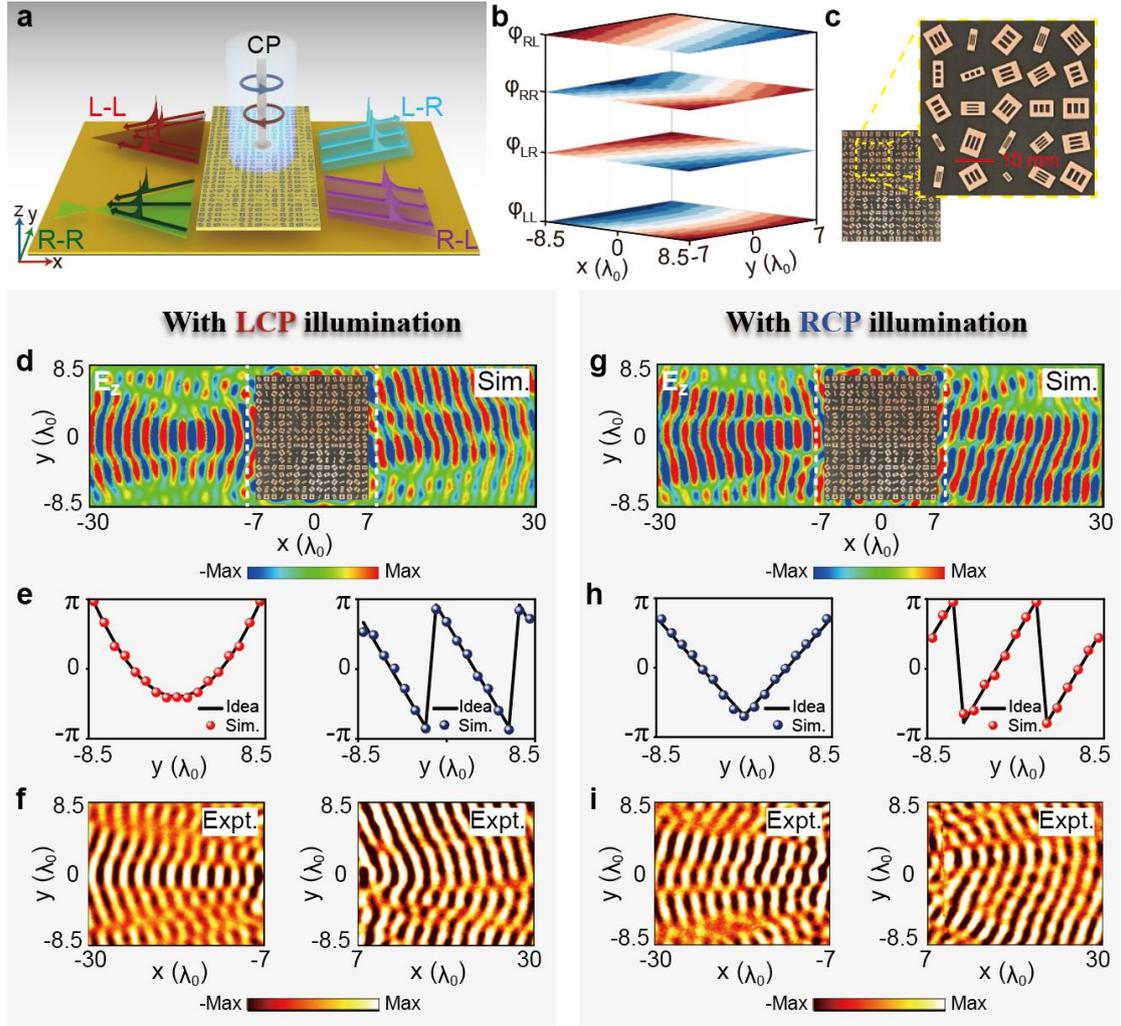

**Fig. 4 Characterizations of the quad-channel SW manipulation metadevice enabling the generation of SWs with distinct functionalities. a** Schematic of the proposed SW metadevice capable of generating four distinct wavefronts: a focused SW, a Bessel-beam SW, and two deflected SWs. **b** Simulated phase profiles of the meta-atoms within the designed metadevice for all four CP transmission channels. **c** Photography of the top section of the metadevice within the fabricated sample. **d-f** Under LCP wave incidence, the simulated near-field Re[$E_z$] patterns in the *xoy* plane (**d**), along with the phase profiles of generated SWs along the white lines (**e**), and the corresponding measured results (**f**). **g-i** When flipping the incident state to RCP incidence, the simulated near-field Re[$E_z$] patterns in the *xoy* plane (**g**), along with the phase profiles of generated SWs along the white lines (**h**), and the corresponding measured results (**i**). Here, the working frequency is 10 GHz.

Based on the phase distributions $\phi_{LL}$, $\phi_{RL}$, $\phi_{LR}$, and $\phi_{RR}$ prescribed by Eq. (3), we designed and fabricated the top-layer metasurface for the second metadevice (see Fig.

S3 in SI for details), a photograph of which is shown in Fig. 4c. Figure 4b presents the simulated phase distribution φ(*x*,*y*) of the transmitted waves from the top-layer metasurface for all four channels, which is also in excellent agreement with Eq. (3), confirming that the desired phase profiles are successfully realized. Figures 4d and 4g depict the simulated Re[$E_z$] field distributions on the plane $z$ = 1 mm at 10 GHz under orthogonal CP incidences. Under LCP incidence, the space wave is efficiently converted into SWs propagating along the left and right plasmonic guiding regions, exhibiting a focusing wavefront and a deflected wavefront, respectively. In the case of RCP incidence, the space wave is similarly converted into propagating SWs but with distinct wavefronts: a focusing wavefront and a deflected wavefront of a different direction, respectively. The calculated conversion efficiencies of this metadevice are 24.3%, 21%, 21.5% and 20.4% for L-L, L-R, R-L and R-R channels, respectively.

We note that although the simulated Bessel and focusing SW wavefronts appear somewhat similar, they are physically distinct. To verify this, we extracted the phase distributions of the excited SWs along the white dashed lines marked in Figs. 4d and 4g, as plotted in Figs. 4e and 4h. The phase profiles for the two functionalities are clearly different: the phase distribution for the focusing wavefront follows a parabolic shape, whereas that for the Bessel wavefront exhibits a pair of symmetric polygonal lines. Using a measurement setup similar to that shown in Fig. 3c, Figs. 4f and 4i show the measured near-field distributions on the detection plane $z$ = 1 mm at 10 GHz under the two orthogonal CP incidences. Clearly, the experimental results agree well with the numerical simulations, further validating the feasibility of the proposed full-channel SW wavefront manipulation strategy. The measured focal length of the focused SWs is approximately $F$ = 180 mm, which is also in good agreement with both the theoretical prediction ($F$ = 185 mm) and full-wave simulation ($F$ = 189 mm).

It should be emphasized that the proposed wavefront manipulation scheme for SWs in all four CP channels represents a general, frequency-independent approach. Extending it to higher frequencies would, however, necessitate advances in high-precision lithography. In addition, the lossy metallic plasmonic resonators used in the present work should be replaced with chiral all-dielectric nanoantennas, and the microwave plasmonic metal should be substituted by a higher-frequency plasmonic structure that can efficiently support SWs.

## 4. Conclusion

In conclusion, we propose a general criterion for constructing metasurfaces that activate all CP channels to enable on-chip multi-channel wavefront manipulation of SWs. The proposed metadevice consists of a top-layer transmission-type metasurface, which generates four independent phase profiles via four CP channels, and a bottom-layer plasmonic guiding region capable of exciting and sustaining the propagation of the generated SWs. The top-layer metasurface is designed by synthesizing chirality-assisted phase, propagation phase, and geometric phase, which allows all elements of the Jones matrix to be decoupled and modulated independently. To validate the design, two metadevices were fabricated and experimentally characterized for on-chip multi-channel SW wavefront shaping. The simulated total efficiencies for these two devices reach approximately 47.4% (including 24.8% for L-L channel and 22.6% for L-R channel) under LCP incidence and 46.0% (including 19.5% for R-L channel and 26.5% for R-R channel) under RCP incidence. The proposed metasurface thus provides a new route for multi-channel SW wavefront manipulation, showing promising potential in high-capacity on-chip communications and integrated photonic systems. This strategy can also be extended to other frequency regimes using appropriately designed chiral meta-structures.

**Acknowledgements:** This work was supported by the Natural Science Foundation of Zhejiang Province (No. LZ25A040002), National Natural Science Foundation of China (Nos. 62575263)

**Author contributions：** S.L., S. S., and L. Z. conceived the idea. M.K., Y.Y., and C.L. conducted the numerical simulations and theoretical analysis. J.L. conducted the experiments. S.L., Y.Y., K.Z., S.S., and L.Z. wrote the manuscript. X. L., H.Z., J. D., J. Y., L. S., and B.Y. participated in the discussion. S. L., K.Z., S.S., and L.Z. supervised the project. All authors participated in the data analysis and read the manuscript.

**Conflict of interest:** The authors declare no competing interests.

**Data availability:** The datasets generated during and/or analyzed during the current study are available from the authors upon reasonable request.

# References

[1] Maier, S. A., *Plasmonics: Fundamentals and Applications.* Springer, New York, 2007.

[2] Liu, J. F. et al. Spin‐controlled reconfigurable excitations of spoof surface plasmon polaritons by a compact structure, *Laser Photon. Rev.* **17**, 2200257 (2023).

[3] Li, S. Q. et al. Localized spoof surface plasmon skyrmions excited by guided waves, *Adv. Funct. Mater.* **33**, 2305789 (2023).

[4] Anglhuber, S. et al. Spacetime imaging of group and phase velocities of terahertz surface plasmon polaritons in graphene, *Nano Lett.* **25**, 2125 (2025).

[5] Deng, W. T. et al. On-chip polarization- and frequency-division demultiplexing for multidimensional terahertz communication, *Laser Photon. Rev.* **16**, 2200136 (2022).

[6] Yang, Y. H. et al. Terahertz topological photonics for on-chip communication, *Nat. Photonics* **14**, 446 (2020).

[7] Kim, T. et al. Terahertz metamaterial on-chip sensing platform for live cancer cell microenvironment analysis, *Chem. Eng. J.* **509**, 161370 (2025).

[8] Gao, X. X. et al. Programmable surface plasmonic neural networks for microwave detection and processing, *Nat. Electron.* **6**, 319 (2023).

[9] Garcia-Vidal, F. J. et al. Spoof surface plasmon photonics, *Rev. Mod. Phys.* **94**, 025004 (2022).

[10] Nauman, A. et al. Topologically engineered strain redistribution in elastomeric substrates for dually tunable anisotropic plasmomechanical responses, *ACS Appl. Mater. Interfaces* **16**, 6337 (2024).

[11] Wang, W. H. et al. On-chip topological beamformer for multi-link terahertz 6G to XG wireless, *Nature* **632**, 522 (2024).

[12] López-Tejeira, F. et al. Efficient unidirectional nanoslit couplers for surface plasmons, *Nat. Phys.* **3**, 324 (2007).

[13] Wang, J. S. et al. Radiation dynamics and manipulation of extreme terahertz surface wave on a metal wire, *Laser Photon. Rev.* **19**, 2400954 (2025).

[14] Kushiyama, Y. et al. Experimental verification of spoof surface plasmons in wire metamaterials. *Opt. Express* **20**, 18238 (2012).

[15] Huang, Y. et al. HR-Si prism coupled tightly confined spoof surface plasmon polaritons mode for terahertz sensing, *Opt. Express* **27**, 34067 (2019).

[16] Hohenau, A. et al. Dielectric optical elements for surface plasmons. *Opt. Lett.* **30**, 893 (2005).

[17] Devaux, E., Laluet, J.-Y. & Stein, B., Refractive micro-optical elements for surface plasmons: from classical to gradient index optics, *Opt. Express* **18**, 20610 (2010).

[18] Yu, N. F. et al. Light propagation with phase discontinuities: generalized laws of reflection and refraction, *Science* **334**, 333 (2011).

[19] Sun, S. L. et al. Gradient-index meta-surfaces as a bridge linking propagating waves and surface waves, *Nat. Mater.* **11**, 426 (2012).

[20] Sun, S. L. et al. Electromagnetic metasurfaces: physics and applications. *Adv. Opt. Photon.* **11**, 380-479 (2019).

[21] Sun, W. J. et al. High-efficiency surface plasmon meta-couplers: concept and microwave-regime realizations. *Light Sci. Appl.* **5**, e16003 (2016).

[22] Zheng, G. X. et al. Metasurface holograms reaching 80% efficiency. *Nat. Nanotechnol.* **10**, 308–312 (2015).

[23] Cui, T. J. et al. Coding metamaterials, digital metamaterials and programmable metamaterials. *Light Sci. Appl.* **3**, e218 (2014).

[24] Yuan, L. Y. et al. Independent phase modulation for quadruplex polarization channels enabled by chirality-assisted geometric-phase metasurfaces. *Nat. Commun.* **11**, 4186 (2020).

[25] Li, S. Q. et al. Unidirectional guided-wave-driven metasurfaces for arbitrary wavefront control. *Nat. Commun.* **15**, 5992 (2024).

[26] Yuan, L. Y. et al. Reaching the efficiency limit of arbitrary polarization transformation with non-orthogonal metasurfaces. *Nat. Commun.* **15**, 6682 (2024).

[27] Xu, H. X. et al. Interference-assisted kaleidoscopic meta-plexer for arbitrary spin-wavefront manipulation. *Light Sci. Appl.* **8**, 3 (2019).

[28] Sun, S. L. et al. High-efficiency broadband anomalous reflection by gradient meta-surfaces. *Nano Lett.* **12**, 6223–6229 (2012).

[29] Wang, S. M. et al. Broadband achromatic optical metasurface devices. *Nat. Commun.* **8**, 187 (2017).

[30] Li, S. Q. et al. Energy-controllable manipulation on surface waves and propagating waves by bifunctional metasurfaces. *Nanophotonics* **15**, e70005 (2026).

[31] Bomzon, Z. E. et al. Space-variant Pancharatnam-Berry phase optical elements with computer-generated subwavelength gratings. *Opt. Lett.* **27**, 1141 (2002).

[32] Jia, M. et al. Efficient manipulations of circularly polarized terahertz waves with transmissive metasurfaces. *Light Sci. Appl.* **8**, 16 (2019).

[33] Khorasaninejad, M. et al. Metalenses: versatile multifunctional photonic components. *Science* **358**, eaam8100 (2017).

[34] Ding, X. M. et al. Ultrathin Pancharatnam-Berry metasurface with maximal cross-polarization efficiency. *Adv. Mater.* **27**, 1195–1200 (2015).

[35] Xu, H. X. et al. Broadband vortex beam generation using multimode Pancharatnam–Berry metasurface. *IEEE Trans. Antennas Propag.* **65**, 7378–7382 (2017).

[36] Maguid, E. et al. Multifunctional interleaved geometric-phase dielectric metasurfaces. *Light Sci. Appl.* **6**, e17027 (2017).

[37] Zhang, L. et al. Spin-controlled multiple pencil beams and vortex beams with different polarizations generated by Pancharatnam-Berry coding metasurfaces. *ACS Appl. Mater. Interface* **9**, 36447–36455 (2017).

[38] Huang, L. L. et al. Helicity dependent directional surface plasmon polariton excitation using a metasurface with interfacial phase discontinuity. *Light Sci. Appl.* **2**, e70 (2013).

[39] Pors, A. et al. Efficient unidirectional polarization-controlled excitation of surface plasmon polaritons. *Light Sci. Appl.* **3**, e197 (2014).

[40] Duan, J. W. et al. High-efficiency chirality-modulated spoof surface plasmon meta-coupler. *Sci. Rep.* **7**, 1354 (2017).

[41] Li, S. Q. et al. Helicity-delinked manipulations on surface waves and propagating waves by metasurfaces. *Nanophotonics* **9**, 3473 (2020).

[42] Wang, Z. et al. Excite spoof surface plasmons with tailored wavefronts using high-efficiency terahertz metasurfaces. *Adv. Sci.* **7**, 2000982 (2020).

[43] Dai, C. H. et al. Multiplexing near-and far-field functionalities with high-efficiency bi-channel metasurfaces. *PhotoniX* **5**, 11 (2024).

[44] Su, H. et al. Active dual-band terahertz surface plasmonic wave excitation using graphene metasurfaces. *Opt. Express* **33**, 14023 (2025).

[45] Tanemura, T. et al. Multiple-wavelength focusing of surface plasmons with a nonperiodic nanoslit coupler. *Nano Lett.* **11**, 2693 (2011).

[46] Wintz, D. et al. Holographic metalens for switchable focusing of surface plasmons. *Nano Lett.* **15**, 3585 (2015).

[47] Jiang, X. H. et al. Geometric phase control of surface plasmons by dipole sources. *Laser Photon. Rev.* **17**, 2200948 (2023).

[48] Wu, M. Z. et al. On-chip multi-channel wavefront manipulation of spoof surface waves with structural heterogeneous metasurfaces. *Laser Photon. Rev.* **19**, e01634 (2025).

[49] Wang, D. P. et al. Bifunctional spoof surface plasmon polariton meta-coupler using anisotropic transmissive metasurface. *Nanophotonics* **11**, 1177 (2022).